# Metaplastic and Energy-Efficient Biocompatible Graphene Artificial Synaptic Transistors for Enhanced Accuracy Neuromorphic Computing

Dmitry Kireev[1,2]*, Samuel Liu[1]*, Harrison Jin[1], T. Patrick Xiao[3], Christopher H. Bennett[3], Deji Akinwande[1,2], and Jean Anne C. Incorvia[1,2]

[1] Department of Electrical and Computer Engineering, The University of Texas at Austin, Austin, TX, USA

[2] Microelectronics Research Center, The University of Texas, Austin, Texas, 78758 USA

[3] Sandia National Laboratories, Albuquerque, New Mexico 87123, USA

* these authors contributed equally

**ABSTRACT**

CMOS-based computing systems that employ the von Neumann architecture are relatively limited when it comes to parallel data storage and processing. In contrast, the human brain is a living computational signal processing unit that operates with extreme parallelism and energy efficiency. Although numerous neuromorphic electronic devices have emerged in the last decade, most of them are rigid or contain materials that are toxic to biological systems. In this work, we report on biocompatible bilayer graphene-based artificial synaptic transistors (BLAST) capable of mimicking synaptic behavior. The BLAST devices leverage a dry ion-selective membrane, enabling long-term potentiation, with ~50 aJ/$\mu m^2$ switching energy efficiency, at least an order of magnitude lower than previous reports on two-dimensional material-based artificial synapses. The devices show unique metaplasticity, a useful feature for generalizable deep neural networks, and we demonstrate that metaplastic BLASTs outperform ideal linear synapses in classic image classification tasks. With switching energy well below the 1 fJ energy estimated per biological synapse, the proposed devices are powerful candidates for bio-interfaced online learning, bridging the gap between artificial and biological neural networks.

**BODY**

As the world becomes more interconnected and data-driven, effective deployment of data-intensive computation methods becomes more critical. Large and complex data structures require constant extrapolation, interpolation, and classification, which are ill-suited for memory-constrained von Neumann architectures[1]. A promising approach for overcoming the power and latency shortfalls of traditional computing is through massively parallel neuromorphic systems[2,3]. A wide variety of devices have been proposed to build such systems, from mature technologies such as metal-oxide[4,5] and phase change memories[6,7] to emerging devices such as electrochemical[8] and magnetic memories[9–12]. Most of these systems, however, employ rigid materials, making them less suited for direct integration with biological matter. Direct interfacing of artificial neuromorphic systems with biological living neurons is a highly ambitious goal, which in the long term may lead to effective brain implants and creation of artificial tissue. Two-dimensional (2D) materials are a promising material class for bioelectronics[13,14] and neuromorphics[15–17] due to their unique electronic properties and atomically thin structure, allowing for imperceptible interfacing with tissue. However, existing 2D-material based neuromorphic systems often employ elements that would induce toxicity when interacting with biological systems, e.g., involving elements like $Li^+$ ionic carriers[18,19] which directly impact the nervous system[20]. As a result, new device innovations are needed for future neuromorphic computing solutions that can directly integrate with biological tissue.

The biocompatible graphene-based artificial synaptic transistors (BLAST) introduced in this work are a combination of two flexible, soft, and biocompatible elements: Nafion and graphene. Nafion plays the role of a solid polymeric electrolyte made of a negatively charged polysulfonated backbone[21] with mobile islands of positively charged water/proton clusters (Fig. 1a). When Nafion is in an environment with a high concentration of protons, channels are formed in the material matrix, allowing for high mobility transport of protons[22]. In contrast, when there is a low concentration of protons, the protonic charge carriers exist in the form of semi-mobile clusters[23]. When a current pulse is applied through the Nafion, we hypothesize that the positively charged clusters move in the opposite direction of electron current and provide an effective change in the local electrical double layer (EDL) at the Nafion-graphene interface, yielding high-precision conductance states for synaptic operation in artificial neural networks. Altogether, the proposed devices feature favorable synaptic characteristics and energy efficiency down to 50 aJ/$\mu m^2$. We evaluate this behavior through neuromorphic simulations on several classification tasks using a prototypical neural network and show that the BLAST devices feature metaplasticity that allows online learning performance exceeding ideal linear, numerical synapse results.

## Main

In this work, two kinds of BLAST devices were fabricated: macroscale (mBLASTs, 10-100 $mm^2$ in area) and microscale ($\mu$BLASTs, ~400 $\mu m^2$ in area). The mBLASTs were fabricated using 180 $\mu$m thick Nafion-117, adhesive conductive gold tape to form source and drain electrodes, and graphene electronic tattoos[24,25] supported by 200 nm thick PMMA (details in Methods). To make mBLASTs, we leverage the previously developed large-scale graphene electronic tattoos (GETs) that are transferred on top of the Nafion, self-adhering and forming a tight interface between the graphene and Nafion. The water-assisted transfer, though rapid (<10 sec), leads to hydration of the Nafion membrane; hence the used Nafion cannot be considered completely dry, leading to relatively increased mobility of proton clusters[22]. The fabricated devices, as shown in Fig. 1a, are highly flexible and transparent. The average bilayer graphene channel width is 4 mm, and length 3-5 mm, yielding an average area of 15 ± 2 $mm^2$. In a separate experiment, we varied the area of mBLAST devices and recorded the change in properties. The gold electrode used to apply synaptic gate potential through the Nafion typically outsizes the graphene channel itself to ensure that the whole graphene-Nafion interface is used effectively. The device schematics and cross-section can be seen in Fig. 1b-c and additional schematic details can be found in Supplementary Fig. S1. The channel conductance of the synaptic device increases in response to negative current pulses applied through the Nafion membrane and decreases in response to positive current pulses, with the amplitude of conductance change correlating with the amplitude of applied pulse (Fig. 1d). The response is symmetrical, reversible, and scalable.

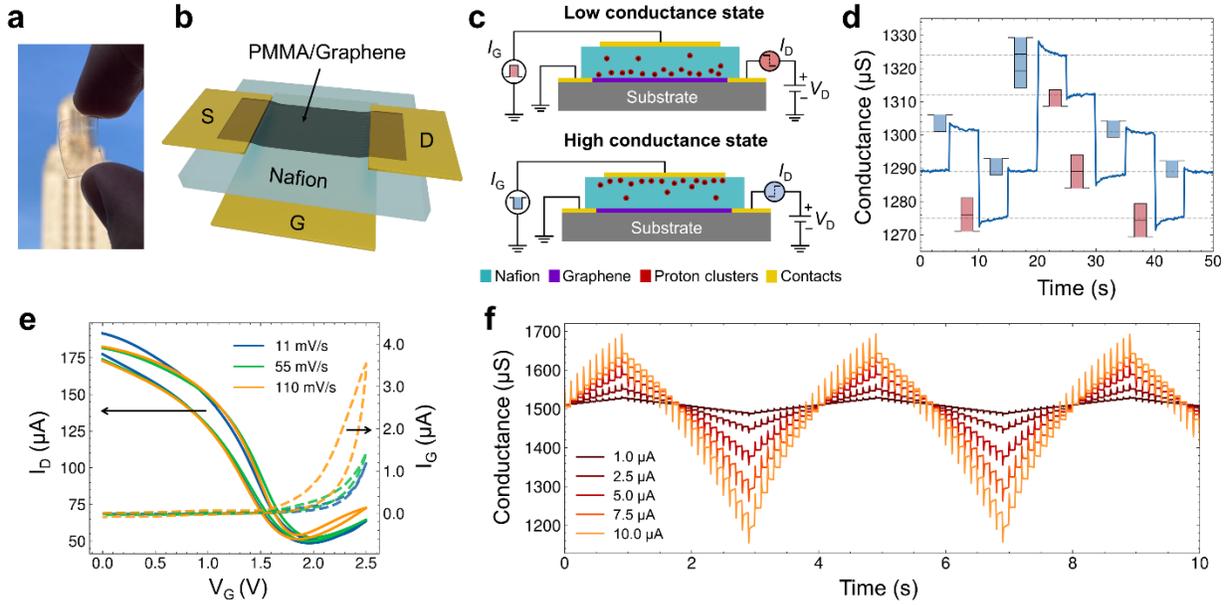

***Figure 1. BLAST configuration and functionality. a,*** *Photograph of a transparent BLAST device.* ***b****, 3D schematic of the BLAST device showing source (S), drain (D), and gate (G) electrodes.* ***c****, Cross-sectional schematic of BLAST device operation in high and low conductance states. Nafion (teal) contains mobile positively charged clusters of protons (dark red) carried through the membrane; source, drain, and gate gold contacts are shown in yellow with source grounded. The pre-synaptic write pulse is labeled $I_G$ and the corresponding post-synaptic read current change is labeled $I_D$. Applied post-synaptic voltage (green) is labeled $V_D$.* ***d****, Measured BLAST conductance vs. time as the gate current is periodically pulsed, showing distinct and repeatable conductance levels. The color/direction and length of the depicted pulses represents the sign and relative amplitude of the applied pulse, respectively.* ***e****, Transfer characteristics with forward and reverse sweep at various sweep rates (11 mV/sec – blue, 55 mV/sec – green, and 110 mV/sec – yellow). Solid lines represent drain-source current; Dashed lines represent gate-source current.* ***f****, performance for positive and then negative trains of 20 write pulses each while gradually increasing current pulse amplitudes from 1 μA (black) to 10 μA (yellow) for 1 ms duration, showing that conductance weights can be modulated by the gate-source current and desired synapse characteristics of high symmetry and linearity.*

We hypothesize that positive (negative) current pulses cause the semi-mobile positively charged proton-concentrated clusters within Nafion to move toward (away) from the graphene interface, shifting the interface capacitance and affecting the electrical double layer, which consecutively affects the charge carriers' density and graphene channel conductance. To corroborate the hypothesis, we gathered transfer curves of the BLAST devices, shown in  by fixing drain voltage $V_D = 0.1$ V, sweeping gate voltage $V_G$ from 0 V to +2.5 V at varying speeds, and measuring both the drain current ($I_D$, solid curves) and gate current ($I_G$, dashed curves). The measurements show that the charge neutrality point (CNP) of the graphene is ~2 V, which means the graphene is highly *p*-doped, which is expected considering the graphene was transferred using wet-etch approach. Faster sweeping rate results in increased hysteresis in gate current $I_G$ and a slight leftward shift on the backward sweep, which is expected behavior for ambipolar transistors[26]. The proton accumulation near the graphene interface is shifting the Fermi energy closer to the CNP, resulting in a decreased conductance. We also observe a slight decrease in minimum $I_D$ (see Fig. S2), which may be an indication that a limited hydrogen adsorption on graphene is happening[27].

Figure 1f shows change in device conductance as a response of cycling pulse trains (positive and negative) for a fixed absolute value current magnitude $I_G$. This is repeated for 5 different current magnitudes, and the result shows that the response is highly reversible and modulated by the strength of the pulse amplitude. To show that the use of graphene is essential in the function of this device, we characterize alternative devices by replacing the active channel layer with another common semiconductor, PEDOT:PSS (see Fig. S3). The control devices show no conductance modulation upon current pulses applied through the Nafion gate, because PEDOT:PSS requires bulk electrochemical reactions with mobile protons to modulate the conductance[8,28]. Since the Nafion is not fully hydrated and there is no proton source at the gate, there are no freely moving protons to deploy[22,23]. Hence, the observed effect is related to the hybrid interface between graphene and Nafion. We hypothesize that the proton clusters migrate toward the graphene-Nafion interface, forming an electrical double layer and either hole- or electron-doping graphene, a similar mechanism to that seen in supercapacitors with carbon-based electrodes[29]. Our observed surface effect requires far fewer protons to accomplish a conductance change than the bulk electrochemical reaction in PEDOT:PSS due to the atomically thin nature of graphene. Coupled with the slow kinetics of the proton cluster movement within the dry Nafion matrix, long-term potentiation can be explained.

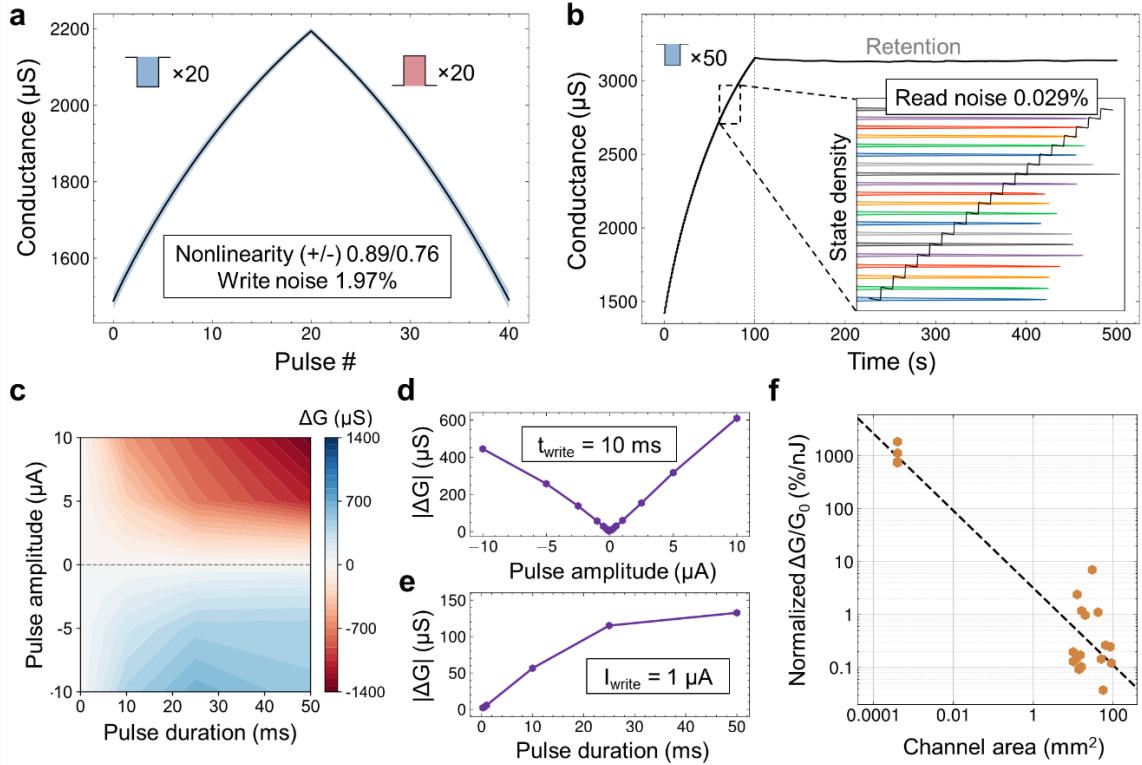

***Figure 2. Performance metrics of the macroscale BLAST devices. a,*** *Conductance per pulse number (20 negative and 20 positive pulses). The black line depicts the mean over 50 measurements, and the blue cloud represents the standard deviation. Nonlinearity values for potentiation and depression were found to be $\delta_P$ = 0.89 and $\delta_D$ = 0.76. Average write noise was found to be $\sigma_{write}$ = 1.97%.* ***b,*** *Stepwise increase in device conductance upon a series of 100 consecutive pulses (each pre-synaptic pulse is 10 μA, 100 ms). The inset shows the state density distribution of 20 states, which do not overlap, indicating extremely low write noise at 0.029% of the dynamic range.* ***c,*** *Color plot of conductance change intensity with varied pre-synaptic pulse amplitude and duration. Supplementary Fig. S6 contains the log-log version of the plot that better represents distribution at lower pulse amplitudes and durations.* ***d,*** *Absolute change in conductance (ΔG) with varied pulse amplitude (pulse duration is constant, 10 ms).* ***e,*** *Absolute change in conductance (ΔG) with varied pulse duration (pulse amplitude is constant, 1 μA).* ***f,*** *Percentage change from initial conductance normalized by energy dissipated as a function of channel area. The black dotted line depicts a line of linear fit with a correlation coefficient of R = -0.93.*

High linearity and symmetry in synapse response are essential figures of merit for efficient backpropagation training of neural networks[30]. The conductance evolution behavior of the mBLAST device is linear and symmetric, indicating the potential for use in online learning applications. This is evaluated by applying a ramp test: a series of repeated positive and negative spikes cycling the device across its dynamic range. Linearity and symmetry are calculated by collapsing 50 repeated ramping cycles into a single ramp (see Fig. 2a), obtaining the mean and standard deviation conductance of the ramps performed.

Non-linearity parameters calculated using previously reported methodology[31] described in Supplementary Note 1 are $\delta_P = 0.89$ and $\delta_D = 0.76$, indicating that performance is approximately linear ($\delta < 1$, where $\delta = 0$ is ideal linearity) and highly symmetric since both directions have the same sign and nearly the same magnitude. As for the number of states achievable, there are 100 explicit conductance states shown in these devices in Fig. 2b. In other experiments, we achieved 300 (see Fig S4) and a maximum of 512 (see Fig S5) specifically defined conduction states, which is more than achieved in comparable device. Ultimately, the actual number of states achievable is directly related to the read and write noise levels. The mBLAST devices feature an average read noise calculated (see Supplementary Note 1) to be 0.029% and write noise of around 1.0% (see Fig. 2a-b), corroborating that a high number of effective states can be functionally achieved in the device. The on/off ratio is also high and spans in the range of 200-500%, which is comparable to or even exceeds results of competing ionic devices[32–34] (see Table S1).

In order to get a complete picture of conductance state control, the relationship between a conductance change ($\Delta G$) per current pulse of different durations and amplitudes is investigated. Mapping of the conductance change as a function of the pulse duration and amplitude (Fig. 2c) shows that $\Delta G$ has an almost linear dependence on the pulse duration and pulse amplitudes. It can be seen (Fig. 2d-e) that $\Delta G$ increases linearly with the increase in either only pulse amplitude or pulse duration. Furthermore, there is an evident asymmetry in the response, which becomes more pronounced at higher pulse duration and amplitude. A long positive pulse at high amplitude will result in a much larger magnitude in conductance change than a long negative pulse of high amplitude. However, at lower pulse duration, the $\Delta G$ is approximately the same magnitude for opposite polarity pulses and generally has a linear dependence with pulse amplitude. The change in conductance for 50 ms long pulses deviates from this linear trend, but this is possibly due to the need for a larger magnitude current pulse to overcome intrinsic energy barriers in ion transport within the dry matrix of Nafion.

Finally, we study how channel area affects device performance by building devices with different channel areas, varied from ~10 mm$^2$ to ~100 mm$^2$. For the experiments, both pulse amplitude and duration are kept constant, and a clear decline in energy normalized conductance change ($\Delta G/G_0$, %/nJ) magnitude is observed (see Fig. 2f) with increasing the device area. Though a linear trend ($R = -0.93$) cannot be concluded due to the appearance of two area-dependent clusters, a general decrease in conductance change as area increases is evident. This is corroborated when only considering devices with channel area larger than 1 mm$^2$ ($R = -0.7$) (see Fig. S7). This means that application of the same gate pulses to smaller devices results in larger effective charge-per-area, resulting in more effective shift in the graphene Fermi level and showing scalability of the BLAST.

To characterize the endurance of a graphene/Nafion synapse, long-term pulse cycling (ramp of 20 pulses with 10 µA amplitude and 1 ms duration) is performed for more than ten million cycles (see Fig. 3a). No device performance degradation was seen for up to 10$^7$ cycles, which is the maximum capacity of our measurement set-up, and not the limit of mBLAST performance. As one can see from the averaged and collapsed ramp (Fig. 3b), the performance for the first 10$^6$ cycles is highly uniform. A more granular representation of the change in ramping characteristics between 10$^6$ and 10$^7$ cycles is shown in Fig. S8. Starting from ~2×10$^6$ cycles, there is a drop in conductance, which saturates at ~4×10$^6$ cycles, and is very uniform afterward. Furthermore, when approaching 10$^7$ cycles, there is a slight increase in the minimum conductance, along with other changes in the response shape of the ramp. It is important to note that this asymmetry happens only at the lowest conduction points when negative pulses are applied through the gate. This means that median electrostatic charge is gradually shifted, reducing the conductance of the graphene device until the CNP is reached (Dirac point, at approx. +1.8 V, see Fig. 1d). A further change in electrostatic charge will further shift the Fermi level into electron-doping regime, increasing the channel conductance. This feature changes device linearity, but due to the similar magnitude of non-linearity and symmetry, conductance changes performed after 10$^7$ pulses would still be effective.

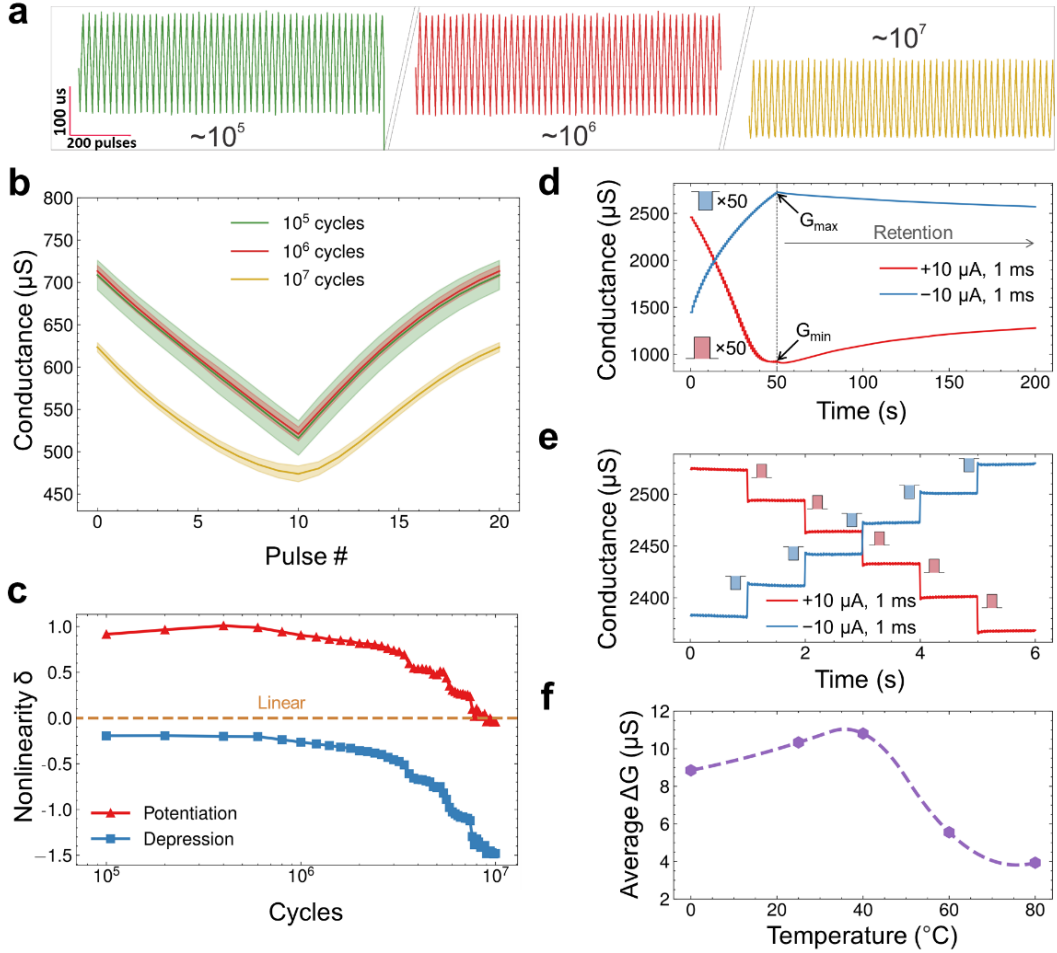

*Figure 3. Endurance and retention characteristics of mBLAST. **a,** Endurance ramp testing using 20 positive and negative pulses per ramp of an mBLAST at $10^5$, $10^6$, and $10^7$ pulses. **b,** Collapsed ramps averaged across $2 \times 10^3$ cycles prior to the cycle number indicated in the legend. The colored cloud corresponds to the write noise of each snapshot. **c,** Nonlinearity parameter $\delta$ calculated as a function of the number of cycles. The orange dashed line depicts perfect linearity. **d,** Conductance relaxation after 50 pulses are delivered to mBLAST to bring the device to minimum and maximum conductance. **e,** Conductance over time with 5 pulses delivered at a one second period. **f,** Average change in conductance as a function of temperature in Celsius.*

Retention time is another essential figure of merit for synaptic transistors and many neuromorphic computation systems[31]. In this work, we report on long retention times exceeding 400 seconds (s), shown in Fig. 2a. Retention time here refers to the time required for 10% decay of the effective conductance range. Some neuromorphic devices report on much longer retention times[8,35]; however, when working with biologically relevant systems, the average frequency of events is <1 Hz. Hence, 400 s retention time is adequate. Another example of this is shown in Fig. 3d, where 50 current pulses are delivered to the device and measurement is continued, allowing the device state to relax. The dominance of short-term potentiation depends on the conductance regime of the device, where changes in conductance are primarily short-term when the conductance is near maximum and minimum. For example, if a positive conductance change is applied when the device is near maximum conductance, the change in conductance will only be retained for several seconds before conductance decays to the value preceding the change. This is an attribute that is common to ionic synaptic transistors of various types[36,37]. In the timescale of 1 pulse/sec (see Fig. 3e), the device exhibits long-term plasticity (no conductance decay is present at the scale). As a result, mBLASTs can be operated in variable conductance ranges depending on the balance of needs between retention and dynamic conductance range.

For effective online learning, individual write pulses must be energy efficient. Devices of varying area from ~10 mm$^2$ to ~100 mm$^2$ were fabricated and tested by applying a set current pulse of 1 μA amplitude, 10 ms duration and recording the change in conductance. The calculation of charge modulation through current spike and energy per spike calculation are given in Supplementary Note 2. The energy dissipation per device area is shown in Fig. 4a and is normalized to reflect the energy required to change the conductance of the device by 1% (chosen to approximate a minimum achievable step due to write noise and read noise). The area-normalized energy dissipation per write pulse is very low, far below 1 fJ, the energy dissipation of a biological synapse. It is evident that there is a weak trend of increasing the energy with smaller device area, with μBLASTs exceeding 1 fJ/μm$^2$ in energy dissipation, but the results still indicate that device scaling still greatly benefits the devices. These results still exceed or match comparable organic and 2D devices at a competitive speed (see Fig. 4b and Table S1).

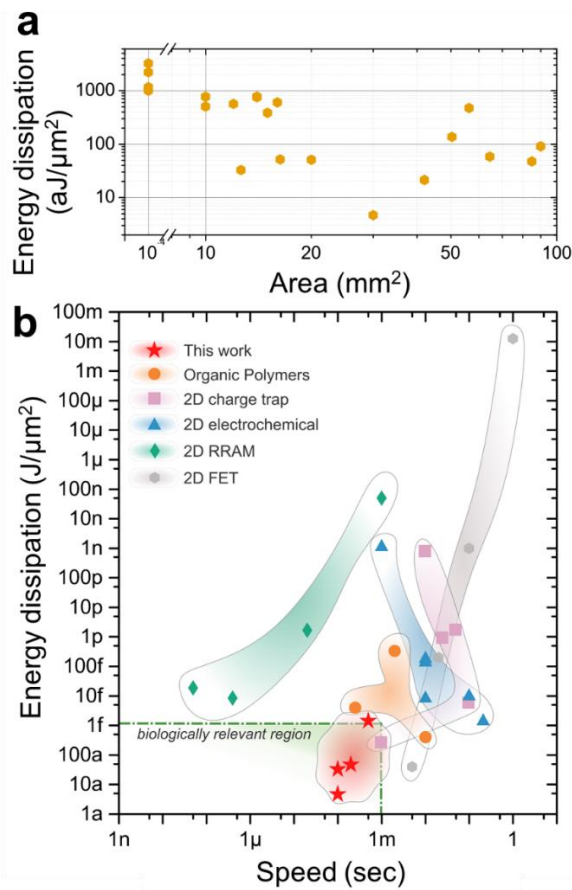

*Figure 4. Comparative performance of BLAST devices. **a**, Energy dissipation normalized by device area and 1% change in conductance as a function of device area. **b**, Benchmarking figure, comparing the BLAST devices (red stars) to other organic polymer-based devices (orange circles) and 2D material based charge trap (purple squares), electrochemical (blue triangles), RRAM (blue rhombus) and FET (gray hexagons) devices. The details and references can be found in* Table S1 *and* Fig. S9.

It is known that Nafion proton conductivity increases with increased temperature[38]. We conducted experiments with an mBLAST device characterized at temperatures in the range of 0°C to 80°C showing marked differences in device performance. As seen in Fig. 3f, the average performance of the synaptic transistor in terms of ΔG improves when temperature is raised from 0°C to 40°C (from ~9 µS to ~11 µS), followed by a decrease down to 4 µS at 80°C. Additionally, while there is no significant change in linearity, there is an increase in write noise up to 22% at 80°C, shown in Fig. S10. We associate the decrease in performance with reduction of Nafion hydration due to annealing at temperatures above 40°C. In bioelectronics-related applications, however, these temperatures should never be reached, and device performance is robust at the relevant temperatures.

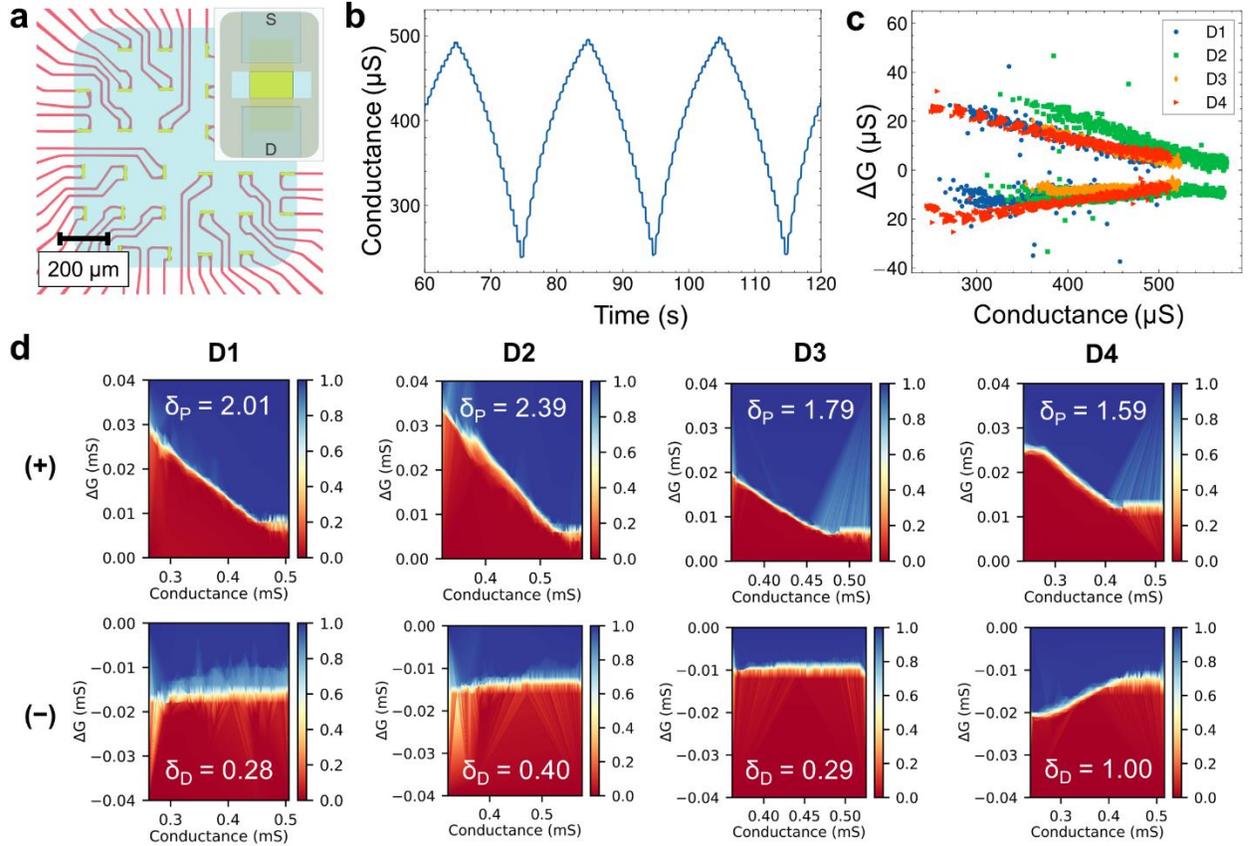

***Figure 5. Performance metrics of the μBLAST devices. a,*** *Schematic of the 32 μBLASTs array, featuring metal feedlines (red), graphene channel (green), passivated with polyimide and covered with Nafion (blue).* ***b,*** *μBLAST conductance cycling per alternating positive and negative trains of 20 write pulses (10 nA magnitude and 5 ms duration).* ***c,*** *Average change in conductance of four nominally identical μBLAST devices showing an overlap in dynamic range and plasticity.* ***d,*** *Heat maps of the cumulative distribution function for the four μBLAST devices, showing the unique metaplastic behavior for positive spikes. On each graph, the color code is normalized from 0 to 1, representing the probability that ΔG is greater than the expected value of ΔG.*

The effect of different drain voltage potential $V_D$ on mBLAST performance was also characterized. We found that modulation of $V_D$ has no effect on the maximum and minimum conductance of the synapse (see Fig. S11) and does not have a clear effect on either linearity or write noise. As a result, the chosen read voltage does not affect device functionality.

Although the aforementioned mBLASTs are biocompatible, feature superior energy efficiency, and have outstanding neuromorphic plasticity characteristics, the devices are still rather large. In order to implement more complex neuromorphic functionalities, the device sizes must be scaled down, which we achieved by building microscale devices (μBLASTs). The μBLASTs were made using UV-lithography with HD-8820 passivated graphene field-effect transistors (GFETs, see Fig. 5a)[39]. On top of the pre-passivated GFET array, a Nafion-117 containing solution was spin-coated and hard-baked, repeating to obtain three layers of Nafion, with the total thickness of ~650 nm. This high thickness ensures there is no short between the top gate and bottom contacts and thinner Nafion could be used. Similar performance synaptic behavior is observed in the μBLASTs (Fig. 5b), with significantly lower currents required to change the conduction state of the device compared to the larger mBLASTs. The performance is notably different compared to the macroscale devices. Firstly, it is important to note that reduction of size yielded an increase in normalized energy dissipation (see Fig. 4a). For devices with channel dimensions 40 μm × 10 μm, energy dissipation is found to be $2.55 \pm 0.39$ fJ/μm$^2$ ($N = 4$) for a single update. This is most likely due to the usage of a liquid-phase Nafion source material, which is not chemically prepared like thicker Nafion films[40]. It is possible that hard-baking the spin-coated Nafion results in less moisture, reducing the density of protonic clusters compared to the preprocessed film. Figure 5c shows sampled ramp data for four 40 μm × 10 μm devices with conductance and $\Delta G$ ranges within similar ranges of operation. This data is then used to construct the $\Delta G$ vs. $G$ lookup table[30] (LUT) graphs presented in Fig. 5d for the four μBLAST devices. The color-coded plots describe the cumulative distribution function likelihood that a given $\Delta G$ is greater than the expected experimental $\Delta G$. Nonlinearity parameters are also presented for each distribution, and it is evident that there was greater non-linearity for the μBLASTs during positive updates and a relatively linear response for the negative updates, indicating that plasticity is asymmetric. While asymmetric update response is typically considered non-ideal for applications like backpropagation training[30,31], we show that such asymmetry of experimental data can yield algorithmic benefits in compute-limited situations.

After analyzing the device characteristics of the mBLASTs and μBLASTs, the experimental conductance update characteristics in the form of LUTs are used to simulate three online learning neuromorphic tasks: (1) Modified National Institute of Standards and Technology (MNIST)[41] handwritten digits recognition, (2) University of California-Irvine Human Activity Recognition (UCI-HAR)[42] dataset for movement classification from biometric sensor data, and (3) Fashion-MNIST[43] clothing article classification. The mBLAST devices were applied on MNIST and UCI-HAR, while the performance of the μBLASTs was evaluated on UCI-HAR and the more complex Fashion-MNIST datasets. Training is accomplished by simulating a small multilayer perceptron with two synaptic weight layers L1 and L2 using CrossSim[44], a physics-rich neuromorphic crossbar simulator. See Methods for more details on the training simulation setup.

Training performance of the mBLASTs on MNIST as well as the application-relevant UCI-HAR dataset was close to ideal, *i.e.,* matching the test accuracy of a comparison network implemented using numeric weights with 64-bit resolution and ideal backpropagation[45] updates (see Fig. 6a-b), reaching ~98% test accuracy on MNIST and ~95% test accuracy on UCI-HAR after 20 epochs. This was a direct result of the desirable characteristics of linearity and symmetry quantified previously (see Fig. 2a), along with low write noise.

Due to the unique non-linear and asymmetric update of the μBLASTs, online training performance of a multilayer perceptron on UCI-HAR learned more slowly than ideal synaptic updates, seen in Fig. 6c. However, after 20 epochs, the difference in classification accuracy between the numeric weights and the BLAST synapses was less than 1%, indicating sufficient synapse expressivity to learn a task at the difficulty of UCI-HAR. Training the network on Fashion-MNIST yielded a surprising result (As evident from Fig. 6d): the μBLAST synapse training performance significantly exceeded that of the ideal numeric weights. The improvement in classification accuracy is more pronounced when utilizing the measured synaptic properties of a single device (D1) compared to when the variation across all four devices (D1-4) is accounted for, indicating that D1 has plasticity characteristics that should be investigated more closely. See Fig. S12 for a learning rate sweep confirming this effect. We hypothesize that this improvement in learning is attributed to the unique shape of the synaptic response (see Fig. 5d), where $\Delta G$ is larger at low conductance values and smaller at high conductance values. In other works from others and ourselves, similar update responses, known as metaplasticity, have been shown to be beneficial for online streamed learning[46,47] and to combat catastrophic forgetting[46].

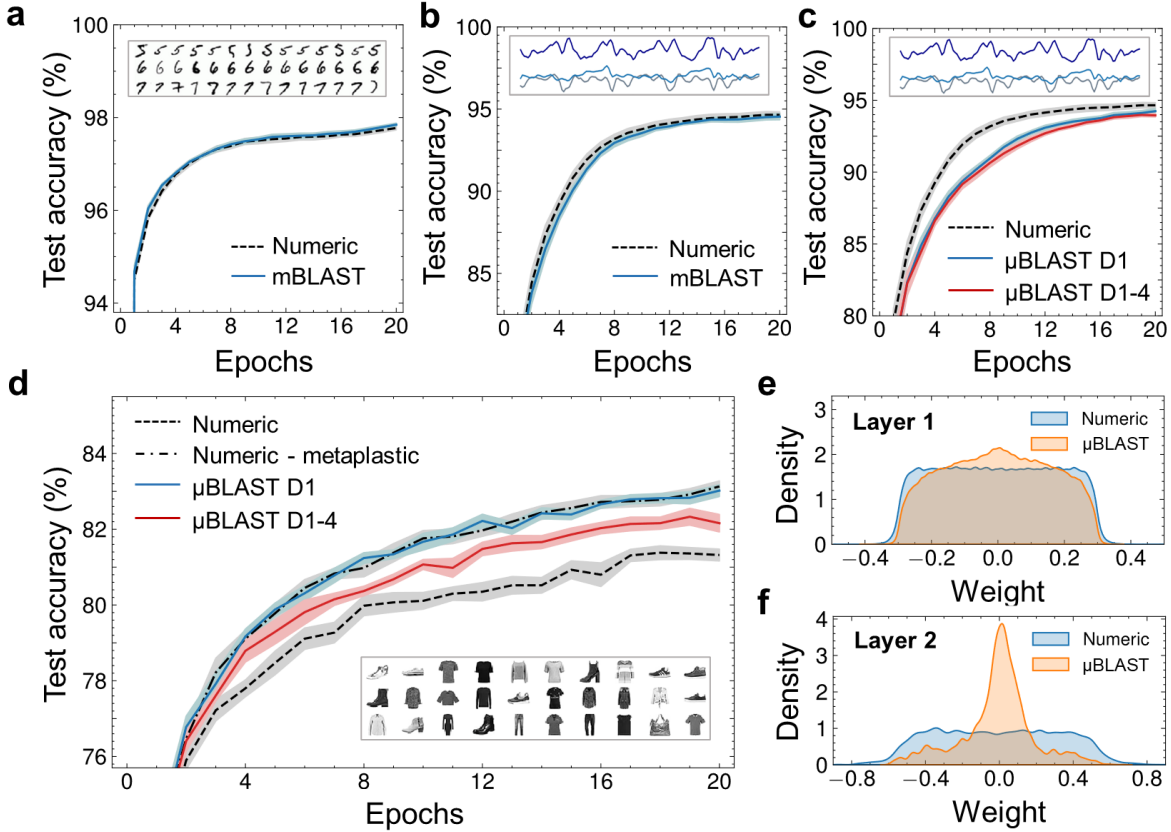

***Figure 6. Neuromorphic crossbar simulation training results. a-b,*** *Online training simulation of multilayer perceptron with mBLAST devices using the experimental data shown in* Fig. 5d *applied on MNIST (**a**), and UCI-HAR (**b**).* ***c-d,*** *Simulation of a μBLAST crossbar using the experimental data shown in Fig. 4 applied on UCI-HAR (**c**), and Fashion-MNIST (**d**). Insets in (a-d) depict the data type for MNIST (handwritten digits) UCI-HAR (human activity signals) and Fashion-MNIST (clothing articles) neuromorphic tasks.* ***e-f,*** *Weight distribution after 20 epochs in layer 1 (**e**), and layer 2 (**f**) of the multilayer perceptron for a crossbar consisting of numeric weights and synapses derived from Device-1. Layer 1 has 235,500 weights and layer 2 has 3010 weights. The detailed dataset is shown in* Fig. S14*.*

To ascertain the functional effect of the metaplastic synapse property, we modified the numeric updates to emulate the LUT of D1. This was done by including a multiplier that depends linearly on the weight value for positive updates while the negative updates were left unchanged. A graphical representation of the update is shown in Fig. S13. The resulting test accuracy, shown by the numeric-metaplastic curve in Fig. 6d, closely matches that of experimental LUT1, showing that it is indeed the metaplastic update property of the μBLAST device that contributes directly to the boost in performance.

To understand how the synaptic metaplasticity affects the high-level properties of the neural network, we examine the final trained weight distributions of two networks: one trained using fully numeric weight updates (without metaplasticity) and one trained using μBLASTs. The representative distributions for the two neural network layers are plotted in in Fig. 6e-f (see Fig. S14 for full dataset, including the weight distributions of a numeric metaplastic synapse network). In layer 1, the numeric weights have a remarkably uniform distribution in values, while for µBLASTs, the weight distribution visibly tapers off away from zero (see Fig. 6e). This effect is more pronounced in layer 2, where there is a prominent peak in weight distribution around zero that is absent from the numeric case. The tendency of the weight distributions to cluster around small absolute values is a natural consequence of the device metaplasticity. The LUTs plot (Fig. 5d) suggests that as the conductance of the device increases, the change in conductance induced by a given pulse is smaller. This has an effect that is equivalent to weight decay regularization[48], which penalizes large weight values to avoid overfitting to the training set and thus improves the ability to generalize to new examples. This effect is also similar to weight normalization[49], a technique inspired by batch normalization[50], both of which are used to train well-regularized neural networks with superior generalizability. The benefit of this regularization was larger for Fashion-MNIST than for the much simpler UCI-HAR classification task, likely because the small multilayer perceptron was relatively under-parameterized for Fashion-MNIST but not for UCI-HAR. Our results indicate that the µBLAST devices can be used to realize hardware-integrated regularization by taking advantage of metaplasticity.

In conclusion, we have demonstrated a novel solid electrolyte gated graphene device with unique artificial synapse behavior. We experimentally revealed that these bio-compatible devices operate at low energy density (<50 aJ/µm$^2$) at >10 kHz speeds, competitive with other 2D materials-based devices and allowing the potential for the interfacing of biological and artificial synapses. Neural network simulations show that the low read and write noise, coupled with the linear and symmetric synaptic characteristics of the mBLASTs allow for near-ideal classification. More importantly, the µBLASTs are demonstrated to be metaplastic synapses that realize weight normalization, a weight regularization algorithm used to train generalizable neural networks. This allows the performance of µBLAST networks to exceed the classification accuracy achieved by ideal linear numeric weights on difficult tasks. These characteristics make BLAST devices promising candidates in the intersection of bioelectronics and neuromorphic computing.

## Methods

**Growth of Graphene/PMMA stack.** The graphene electronic tattoos were fabricated starting with monolayer CVD grown large-scale graphene on copper, purchased from Grolltex. A~2x2 in$^2$ square was then cut out and spin-coated with PMMA (PMMA 950 A4) to yield a ~200 nm thick layer. To do so, the copper/graphene stack must be placed on a silicon wafer with the graphene facing up. Kapton tape is then used to secure all sides of the copper/graphene stack to the wafer such that a watertight seal is formed under the copper to prevent PMMA from leaking underneath the stack. Afterwards, PMMA (950 A4) is spin-coated onto the stack at ~2,500 rpm for 60 seconds. The sample is then baked on a hotplate at 200° C for 15-20 minutes before it is ready for etching.

**Graphene/PMMA stack transfer onto tattoo paper.** The PMMA/graphene/Cu foil is then placed into ammonium persulfate ($(NH_4)_2S_2O_8$, 0.1M) to etch away the copper. The PMMA/graphene film is then cleaned in a series of water washing steps and then transferred onto a temporary tattoo paper. The graphene/PMMA/tattoo paper is then dried and cut into arbitrary desired locations. In order to transfer the graphene/PMMA onto an arbitrary surface, it is first soaked in water for ~5-10 minutes.

**mBLAST device fabrication.** The macroscale BLAST devices were fabricated in multiple steps, by combining Nafion 117, PMMA/graphene, conductive adhesive gold tape, and gold/EVA/PET film. The Gold/EVA/PET film is used to form a tight back-gate contact. The fabrication starts by evaporating ~60-90 nm of gold onto an ethylene vinyl acetate (EVA) film. After preparing the gold EVA gate, it is brought in contact with a piece of Nafion-117 of the desired dimension (typically from 5x5 mm to 15x15 mm). This step is performed on a hotplate (at ~150 °C) for no longer than 15 seconds. Following the application of the Nafion strip, adhesive gold contacts are placed 3-10 mm apart from each other, perpendicularly from the gate, on top of the Nafion to form the source and drain terminals. The GETs (~5-15 cm length and 3-10 mm width) are prepared for transfer onto the BLAST device starting by soaking in DI water for 5-10 minutes. The graphene should show signs of slight separation from the tattoo paper when sufficiently soaked. The GET can then be placed on top of the Nafion and transferred onto the top of the device such that the graphene contacts both gold source-drain contacts and forms a channel across the Nafion.

**μBLAST device fabrication.** CVD grown graphene was covered with a 200 nm thin layer of PMMA for transfer. After etching copper in ammonium persulfate (see above), the graphene was transferred onto the wafer with pre-fabricated Au/Ti (10/50 nm thick) markers. A photoresist was used to protect the graphene channel areas during exposure to oxygen plasma. The stack of 10 nm Ti and 90 nm Au was e-beam assisted evaporated on the wafer through a pre-defined structure of lift-off resists to form source and drain electrical connections. Photostructurable polyimide, HD8820 was used in the last step to form the passivation. Spin-coated at 5000 rpm, exposed at i-line UV light, developed in 0.26% TMAH, and hard baked at 350°C, the polyimide forms a 3 μm thick passivation. The devices were then diced and spin-coated with the liquid Nafion-117 containing solution (Sigma-Aldrich) three times (3000 rpm, 150° C bake for 20 mins), forming a 666.7±28.9 μm thick layer.

**Pulse and synaptic measurements.** The three-terminal devices were measured using high-precision source/measure unit Agilent 2902B. One of the SMUs is used to apply 0.1 V of the drain-source potential, while the gate is used in the current-pulsing mode to apply conductance changes.

**Pulse and relax.** Two pulse train tests were used to characterize conductance change due to periodic pulsing. This was done by sending either 50 or 250 consecutive write current pulses with corresponding widths and amplitudes of 100 μs and 100 μA, 1 ms and 10μA, 10 ms and 1 μA, and 100 ms and 100 nA. All pulse trains were conducted with both positive and negative pulses. The pulses were followed with a 150-200 seconds relax to observe conductance retention.

**Ramp and level.** A pulse ramp experiment ran multiple sequences of a train of 20 negative pulses followed by a train of 20 positive pulses. The level experiment consisted of an arbitrary sequence of negative and positive current pulses of 10, 20, and 30 µA that were pulsed through the gate every 5 seconds at a pulse width of 1 ms. Both tests were used to demonstrate distinct and repeatable conductance levels.

**Temperature tests.** The temperature dependence test measured macroscale device performance at varying temperatures (0°C, 25°C, 40°C, 60°C, and 80°C). The 0° test was performed inside of an insulated Styrofoam container filled with ice. The macro device was lowered onto ice using a petri dish and connected to the Keysight B2902A using wires thin enough as to not disturb the insulating seal. At 25°, 40°, 60°, and 80°, the device was secured to the top of a hotplate using Kapton tape while a thermocouple was used to verify the temperature. The above-mentioned tests were performed at each temperature with a minimum time of 30 minutes after each temperature change to ensure the device was successfully brought to temperature.

**Neuromorphic computing simulations**. Neural network training simulation results and synaptic update lookup tables were generated from experimental data using CrossSim[44]. During a weight update, the ideal backpropagation update is first calculated. To account for synaptic device nonlinearity and write noise, the probabilistic LUT of device updates is queried to sample a conductance change $\Delta G$ that depends on the current value of the conductance G of a synapse device. To model device-to-device process variations, each device in the simulated array is randomly assigned to one of four (three) LUTs, each of which is constructed from a different experimentally characterized µBLAST (mBLAST) device. To represent both positive and negative weights, a single weight value is encoded in the difference in conductance between two paired BLAST devices. This is done for two reasons: (1) computing the difference in analog reduces the required dynamic range of the peripheral circuits, (2) using two devices per weight, rather than a fixed analog offset for all devices, avoids coupling the variability and noise in the offset resistor into all the results, which effectively amplifies the variability. When a weight is updated, the conductance of both devices in the pair are updated. The weights are trained using stochastic gradient descent with a batch size of 1, i.e. training samples were shown one at a time. A two-layer MLP was trained for each task, where a sigmoid activation is used after the first layer and a softmax is used after the second layer. The MNIST and Fashion-MNIST networks use a 785×300 weight matrix for the first layer and a 301×10 weight matrix for the second layer (including bias). The UCI-HAR MLP has a 577×200 matrix and a 201×6 matrix for the first and second layer, respectively.

**Associated Content**

*Supplementary Information* is available and includes comparison table; supplementary notes with calculations of nonlinearity, read and write noise, and energy dissipation; control experiments; additional schematics; detailed datasets and figures.

**Author Information**

*Corresponding Author*

Email: incorvia@austin.utexas.edu

*Author Contributions*

D. K. and S. L. contributed equally to designing the project, carrying out the measurements, analyzing the data, and writing the paper. H. J. measured the devices and analyzed the data. T. P. X. and C. H. B. carried out the neural network simulations and analyzed the data. D. A. guided the experimental setup, measurements, and direction of the project. J. A. C. I. led the team in the design, measurement, analysis, and writing. The manuscript was written with input from all authors and all authors discussed the results.

*Acknowledgments*

This research was primarily supported by the National Science Foundation (NSF) through the Center for Dynamics and Control of Materials: an NSF MRSEC under Cooperative Agreement No. DMR-1720595. This research is sponsored in part by the National Science Foundation under CCF award 1910997. This material is also based upon work supported by the National Science Foundation Graduate Research Fellowship under Grant No. 2021311125. H. J. acknowledges support from the George L. McGonigle Excellence Fund in Electrical Engineering. D.A acknowledges the Office of Naval Research Grant No. N00014-18-1-2706. This research is sponsored in part Sandia National Laboratories. This paper describes objective technical results and analysis. Any subjective views or opinions that might be expressed in this paper do not necessarily represent the views of the U.S. Department of Energy or the United States Government. Sandia National Laboratories is a multimission laboratory managed and operated by NTESS, LLC, a wholly owned subsidiary of Honeywell International Inc., for the U.S. Department of Energy's National Nuclear Security Administration under Contract No. DE-NA0003525.

The work was done at the Texas Nanofabrication Facility supported by NSF Grant No. NNCI- 1542159 and at the Texas Materials Institute (TMI).

*Notes*

The authors declare no competing financial interest.